\documentclass{article}
\usepackage[utf8]{inputenc}
\usepackage[english]{babel}
\usepackage{graphicx}
\usepackage{subcaption}
\usepackage{amssymb}
\usepackage{dirtytalk}
\usepackage{listings}
\usepackage{csquotes}
\usepackage{amsmath}
\usepackage{amsthm}
\usepackage{multirow}
\usepackage{algpseudocode} 
\usepackage{algorithm} 
\usepackage{xcolor}
\usepackage{dirtytalk}
\usepackage{authblk}

\theoremstyle{definition}
\newtheorem{definition}{Definition}[section]

\usepackage[margin=0.9in]{geometry}

\usepackage{comment} 

\usepackage[
backend=biber,
style=numeric,
maxnames=19,
sorting=none,
citestyle=numeric
]{biblatex}
\title{Multi-Species Prey-Predator Dynamics During a Multi-Strain Pandemic}

\author[1*]{Ariel Alexi}
\author[1]{Ariel Rosenfeld}
\author[2]{Teddy Lazebnik}

\affil[1]{Department of Information Science, Bar-Ilan University, Ramat-Gan, Israel}
\affil[2]{Department of Cancer Biology, Cancer Institute, University College London, London, UK}
\affil[*]{Corresponding author: ariel.alexi@live.biu.ac.il}

\date{}

\begin{document}

\maketitle
\begin{abstract}
\noindent
Small and large scale pandemics are a natural phenomenon repeatably appearing throughout history, causing ecological and biological shifts in ecosystems and a wide range of their habitats. These pandemics usually start with a single strain but shortly become multi-strain due to a mutation process of the pathogen causing the epidemic. In this study, we propose a novel eco-epidemiological model that captures multi-species prey-predator dynamics with a multi-strain pandemic. The proposed model extends and combines the Lotka–Volterra prey-predator model and the Susceptible-Infectious-Recovered (SIR) epidemiological model. We investigate the ecosystem's sensitivity and stability during such a multi-strain pandemic through extensive simulation relying on both synthetic cases as well as  two real-world configurations.

Our results are aligned with known ecological and epidemiological findings, thus supporting the adequacy of the proposed model in realistically capturing the complex eco-epidemiological properties of the multi-species multi-strain pandemic dynamics. \\ \\
\noindent
\textbf{keywords:} Lotka–Volterra model, ODE on a graph, numerical simulation, extended SIR model.
\end{abstract}

\section{Introduction}
In nature, a gentle biological and ecological balance is kept in a complex system of plants, animal species, and the environment \cite{bg_1,bg_2,bg_3, sex_ducks_and_rockenroll, choas_prey-predator-mixed-mode, chaos_impact_predator}. In the micro-level of a small spatial location, the ecological system's dynamics is the sum of interactions between a number of animal (and plants) species with their environment and each other. Thus, one can divide these biological interactions into two: animal-environment and animal-animal interactions \cite{aa_interactions}. The first type is mostly stable over time, as changes in such interactions result from a long-term evolution process \cite{evolution}. As such, a good approximation to these dynamics can be associated with the environment's ability to support its inhabitants \cite{live_in}. The latter type is much more complex with multiple ways and strategies animals apply to survive and produce offsprings \cite{offsprings_1,offsprings_2,offsprings_3}. 

This system is highly sensitive and even a small-size event can break this gentle balance and put the ecological system in a long-term course of re-stabilization \cite{ecosystem_1,ecosystem_2}. 
A large catastrophic event can result in fatal outcomes such as species extinction \cite{large_event}, partially ruined food-chains \cite{food_chains_1,food_chains_2,food_chains_3}, and large-scale economic damages for the human population supported by this ecosystem \cite{economic_1,economic_2}. In history, experts recorded many types and occasions of such events, ranging from large-scale fires to extreme weather changes \cite{bad_events_1,bad_events_2,bad_events_3}. A dominant type of event that repeats itself over time and locations is pandemics \cite{pandemic_important,pandemic_duration,pandemic_review}. For example, the influenza virus, a member of the Orthomyxoviridae family, infects multiple species worldwide, including poultry, swine, humans, horses, seals, and other animals \cite{virus_intro_1,virus_intro_2,virus_intro_3}. 

Currently, our understanding of multi-species pandemic is limited due to the complexity of detecting it on time, gathering relevant data, and influencing its course \cite{hard_msms_1,hard_msms_2,hard_msms_3,hard_msms_4}. However, the study of interacting species has gained popularity in the last decades, constantly revealing new insights about the biological dynamics around us and providing a cornerstone for a broad spectrum of technological developments \cite{tech_1,tech_2,tech_3,tech_4}. Indeed, a particular interest is provided to the study of epidemiology to understand the spread of infectious diseases with the goal to determine pandemic intervention policies to possibly eradicate them \cite{pip_1,pip_2,pip_3,pip_4,pip_5,pip_6, building_sir_ariel}. In a complementary manner, the research of prey-predator dynamics has been widely extended with models increasing in complexity and scope which are, presumably, capable of better representation of the dynamics found in nature \cite{nature_1,nature_2}. 
As such, mathematical models and computer simulations are shown to be powerful tools to understand the biological and ecological dynamics of pandemic spread \cite{powerful_tool_1,powerful_tool_2,powerful_tool_3,powerful_tool_4, alexi2022trade, spatio_temporal_airborne_andemic_ariel, choas_Prey-predator_model}. 

A large body of work aims to extend the simple Susceptible-Infectious-Recovered (SIR) model that takes the form \cite{first_sir}:
\begin{equation}
    \begin{array}{l}
         \frac{dS}{dt} = -\beta S(t) I(t), \; \frac{dI}{dt} = \beta S(t) I(t) - \gamma I(t), \; \frac{dR}{dt} = \gamma I(t),
    \end{array}
\end{equation}
where \(S, I,\) and \(R\) are the groups of susceptible, infected, and recovered individuals, respectively, and the average infection rate and the average recovery rate are donated by \(\beta \in \mathbb{R}^+\) and \(\gamma \in \mathbb{R}^+\), respectively. The SIR model assumes the population is well-mixed (i.e., the probability two individuals interact at any point in time is uniformly distributed) and that \(S + I + R = N \in \mathbb{N}\) such that \(N\) is the constant, over time, population's size. As the SIR model is shown to be too simplistic to capture realistic pandemic scenarios \cite{review_1,review_2,review_3}, multiple extensions were proposed to improve it \cite{multi_populations_1,multi_populations_2,multi_populations_3,multi_populations_4,multi_strain_3}. For instance, \cite{sir_example_1} used the SIR model to capture the pandemic spread in a fish population, showing the model is able to well capture and predict the pandemic spread dynamics. \cite{sir_example_2} analyzed an SEIR (E stands for the Exposed status) epidemiological model with healthcare treatment pandemic intervention policy for Ebola in humans. \cite{sir_example_3} reviewed several mathematical modeling attempts for spatial-temporal transmission dynamics of influenza. In particular, they show that spatio-temporal stochastic SIR models are suitable to well approximate the average reproduction number of the swine flu based on historical data. 
More advanced epidemiological models take into consideration multi-strain dynamics where there is more than one pandemic in parallel \cite{teddy_multi_strain}. \cite{multi_strain_2} have studied a class of multi-strain deterministic epidemic models that extend the SIR model in which cross-immunity varies with the genetic distance between strains. The authors show that for low maximal values of cross-immunity, all strains play a critical role in the course of the dynamics and tend to chaos. However, for the complementary case, the system has both chaotic and stable phases during the dynamics. \cite{multi_strain_1} studied a multi-scale immuno-epidemiological model of influenza viruses including direct and environmental transmission, extending the SIR model to allow two time-since-infection structural variables. \cite{multi_populations_3} examine a spatio-temporal model for the disease, extending the SIR model by taking into consideration a population living on two or more patches between any pair of which migration is allowed. They analyzed the influence of a pulse vaccination strategy, concluding conditions for eradicating the pandemic.

In a similar manner, researchers investigated the prey-predator dynamics from a bio-mathematical perspective. Most of the prey-predator models are based on the Lotka-Volterra model which takes the form \cite{lotka_volterra}:
\begin{equation}
    \begin{array}{l}
         \frac{dx(t)}{dt} = ax(t) - by(t)x(t), \; \frac{dy(t)}{dt} = cx(t)y(t) - dy(t),
    \end{array}
\end{equation}
where \(x(t)\) and \(y(t)\) are the prey and predator population sizes over time, respectively. \(a \in \mathbb{R}^+ \) is the natural growth rate of the prey population supported by the environment, \(b \in \mathbb{R}^+ \) is the proton of the prey population that consumed by the predator population, \(c \in \mathbb{R}^+ \) is the rate of resources available for the predator population to grow due to consumption of the prey population, and \(d \in \mathbb{R}^+ \) is the natural decay rate of the predator population. These models are based on two assumptions: a) the habitat for the prey is assumed to be unlimited so that in absence of predators the prey will reproduce exponentially, and b) the predators survive only on the prey, and in the absence of food, their number will decrease exponentially. This model and its extensions are well studied \cite{lv_example_1,lv_example_2,lv_example_3,lv_example_4,lv_example_5}.

Several attempts of merging these two models have been investigated \cite{ls_pandemic_general_1,ls_pandemic_general_2,ls_pandemic_general_3}. \cite{lv_pandemic_2} developed and analyzed a predator-prey model, where both species are subjected to parasitism. They show that in the case where the uninfected predator cannot survive only on uninfected prey, the parasitization could lead to the persistence of the predator provided a certain threshold of transmission is surpassed. \cite{lv_pandemic_3} analyzed a two-species prey-predator model with the SIS epidemiological model where predators have an alternative food source rather than the prey. \cite{lv_pandemic_4} also investigate a two-species prey-predator model with the SIS epidemiological model, proposing a stochastic version of it and a numerical scheme to solve the model efficiently. 
Common to these works is the focus on both single-strain pandemics and only two species. To the best of our knowledge, no model that combines multi-strain epidemiological dynamics with a multi prey-predictor network has been proposed yet. 

\begin{figure}[!ht]
    \centering
    \includegraphics[width=0.99\textwidth]{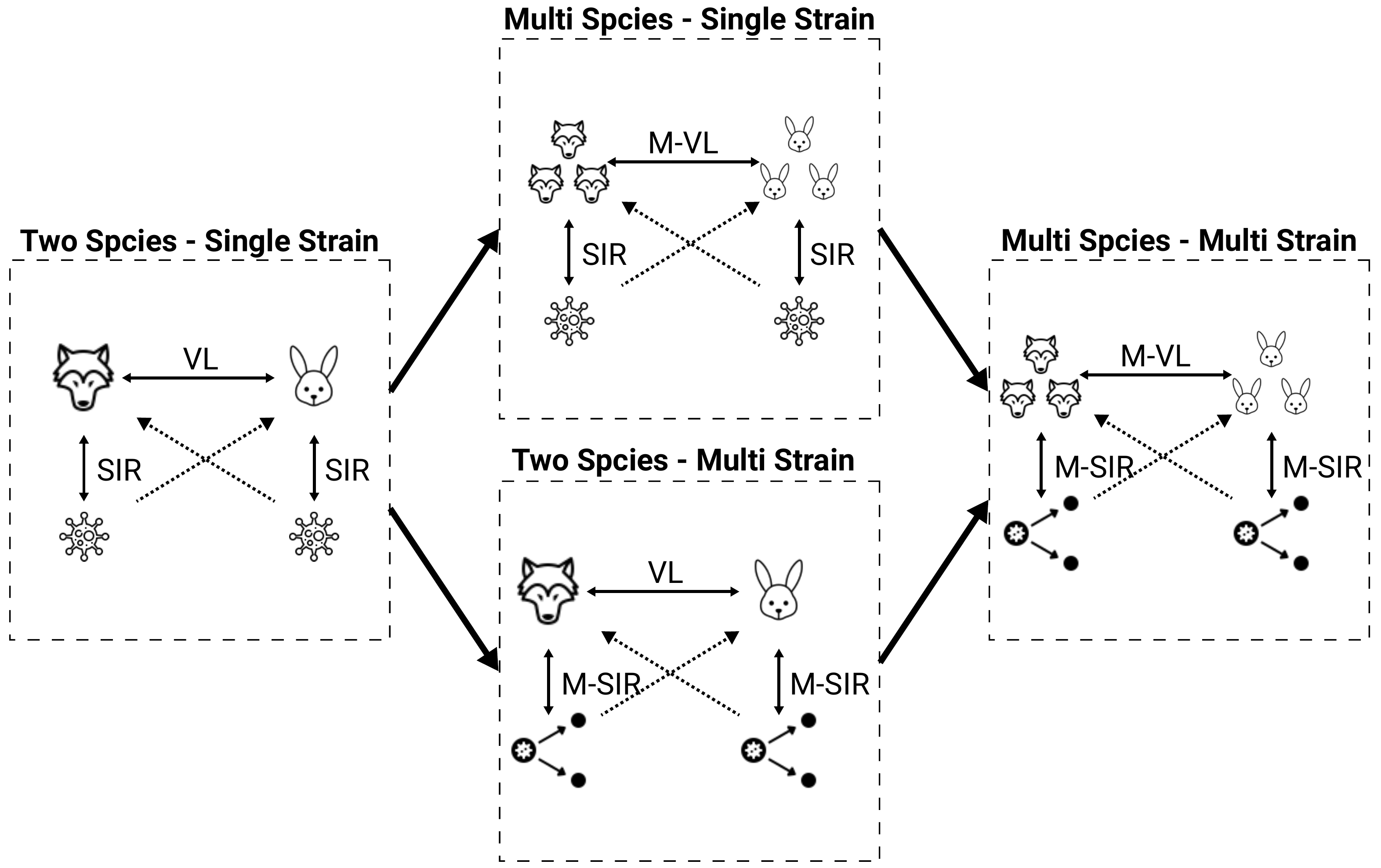}
    \caption{A schematic view of the model's structure and the connections between them.}
    \label{fig:model_scheme}
\end{figure}

In this work, we propose a novel multi-strain with multi-species (MSMS) model for studying the spread of infectious diseases in a more realistic, complex ecosystem. A schematic view of the two possible extensions of two species with a single strain and their merge into an MSMS model is provided in Fig. \ref{fig:model_scheme}. 

The remaining paper is organized as follows: Section \ref{sec:model} describes the proposed model's mathematical formalization followed by a computer simulation implementation. Section \ref{sec:simulator} outlines the implementation of the proposed model as a computer simulator. Section \ref{sec:analysis_numerical} provides a comprehensive evaluation of the proposed model using synthetic and real-world setups. Finally, Section \ref{sec:discussion} provides a discussion on the model's benefits and limitations followed by conclusion remarks and suggestions for future work.  

\section{Model Definition}
\label{sec:model}
In order to capture the ecological-epidemiological dynamics, we use a system of ordinary differential equations (ODEs). Intuitively, we combine the multi-strain pandemic model proposed by \cite{teddy_multi_strain} with a multi-species Lotka-Volterra model proposed by \cite{lv_pandemic_3}. On top of that, we further extend the multi-strain pandemic model to include cross-spices infection and infection's exposition phase. 

\subsection{Single Specie-level dynamics}
\label{section:inner-species}
For each specie in the set of all species in the system, the multi-strain epidemiological model considers a population \(\mathbb{P}_i\) for the \(i_{th}\) specie. We assume a pandemic for specie \(i\) has \(M_i := \{1, \dots, m_i\}\) strains. Each individual in the population, \(p_i^j \in \mathbb{P}_i\), is associated with one of five epidemiological states with respect to each strain: susceptible \((S)\), exposed \((E)\), infected \((I)\), recovered \((R)\), and dead \((D)\). Thus, the epidemiological state of an individual can be represented by a vector \(\eta_i \in \mathbb{R}^{|M_i| \times 5}\). Moreover, as it is assumed that an individual can not be infected or exposed to more than one strain at the same time and since once an individual is dead due to one strain it is dead, the individual's epidemiological state can be reduced to a set of strain one recovered from, \(j \in P(M_i)\), and the current infectious strain, \(k \in P(M_i)\).

\begin{figure}[!h]
\center{\includegraphics[width=0.99\textwidth]{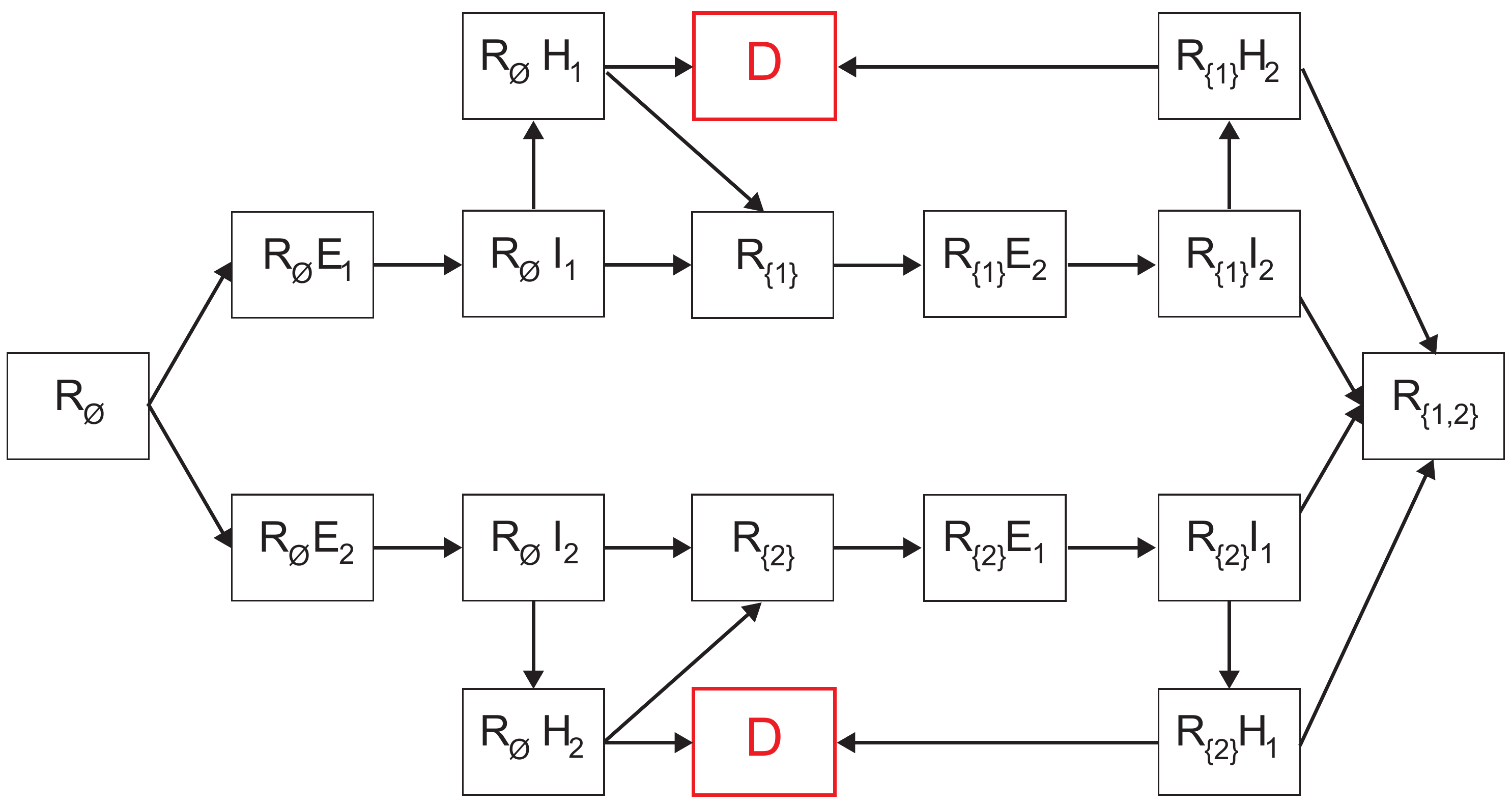}}
\caption{Schematic view of transition between disease stages, shown for \(\|M\|=2\).}
\label{fig:individual_flow}
\end{figure}

Therefore, each individual belongs to one of five groups: 1) Infectious with strain \(k \in M_i\) and a history of recoveries \(J \in P(M_i)\) (i.e., the power set of the strain and its strain set) represented by \(R_J I_k^i\); 2) Exposed with strain \(k \in M_i\) and a history of recoveries \(J \in P(M_i)\) represented by \(R_J E_k^i\); 3) Recovered with a history \(J \in P(M)\) represented by \(R_J^i\); and 4) Dead \((D^i)\). Of note, for \(J = \emptyset\), \(R_J \equiv S\) is the susceptible epidemiological state. A schematic view of the transition between the stages of the disease for an individual for two strains (i.e., \(\|M\|=2\)) is shown in Fig. \ref{fig:individual_flow}. 

Individuals in the Recovered (\(R_J\)) group have immunity to strains \(k \in J\) and are susceptible to infection by strains \(M_i \backslash J\). When an individual in this group is exposed to strain \(k \in M \backslash J\), the individual is transferred to the Exposed with strain \(k\) with a history of recoveries group \(J\) (\(R_J E_k\)) at a rate \(\beta_{J, k}\). The individual stays in this group \(\psi_{J, k}\) time stamps, after which the individual is transferred to the Infected group of the same strain \(k\) with the same recovery history \(J\) marked by (\(R_J I_k\)). The individuals stay in this group \(\gamma_{J, k}\) time stamps, after which the individuals are transferred to the Recovered group (\(R_{J \cup \{k\}}\)) or the Dead group \((D)\) at rate \(1 - \xi_{J, k}\) and \(\xi_{J, k}\), respectively. The recovered individuals are again healthy, no longer contagious, and immune from future infection from the same strain \(k\). 

Formally, for the \(i_{th}\) specie, the multi-strain epidemiological model takes the following form: First, in Eq. (\ref{eq:1}), \(\frac{dR^i_J(t)}{dt}\) is the dynamic number of individuals who have recovered from a group of strains \(J \in P(M_i)\) over time. It is affected by the following two terms. First, for each strain \(k \in J\), an individual who has recovered from group \(J \backslash \{k\}\) of strains and is infected with strain \(k\), recovers at rate \(\gamma^i_{J \backslash \{k\}, k}\) with a probability of \(1 - \xi^i_{J \backslash \{k\}, k}\). Second, individuals infected by strain \(k\) with rate \(\beta^i_{J, k}\). These individuals can be infected by any individual with a strain \(k\) who has recovered from any group \(L\) of strains so that \(k \not\in L\). 

\begin{equation}
    \begin{array}{l}
    \frac{dR^i_J(t)}{dt} = \sum_{l \in J} \gamma^i_{J \backslash \{l\},l} (1-\xi^i_{J \backslash \{i\}, i}) R^i_{J \backslash \{i\}} I_{i}(t) - \sum_{l \in M_i \backslash J} \beta^i_{J, l} R^i_J(t) \sum_{L \in P(M) , c \not\in L} R_{L}I_{c}^i(t) .
    \end{array}
    \label{eq:1}
\end{equation}

Second, in Eq. (\ref{eq:2}), \(\frac{dR_J E_k(t)}{dt}\) is the dynamic number of individuals who have recovered from a group of strains \(J\) and are exposed to a strain \(k\) over time. It is affected by the following two terms. First, individuals infected by strain \(k\) with rate \(\beta^i_{J,k}\). These individuals can be infected by any individual with a strain \(k\) who has recovered from any group \(L\) of strains so that \(k \not\in L\). Second, individuals exposed to strain \(k\) who become infected at rate \(\phi^i_{J,k}\). 
\begin{equation} 
    \begin{array}{l}
        \frac{dR_J E_k^i(t)}{dt} = \sum_{k \in M_i \backslash J} \beta^i_{J,i} R_J^i(t) \sum_{L \in P(M_i) , k \not\in L} R_{L}I_{k}^i(t) - \psi^i_{J, k} R_{J}E_{k}^i(t).
    \end{array}
    \label{eq:2} 
\end{equation}

Third, in Eq. (\ref{eq:3}), \(\frac{dR_J I_k^i(t)}{dt}\) is the dynamic number of individuals who have recovered from a group of strains \(J\) and are infected with strain \(k\) over time. It is affected by the following two terms. First, individuals exposed to strain \(k\) with a history of \(J\) who become infected with strain \(k\), at rate \(\phi^i_{J, k}\). Second, individuals infected with strain \(k\) who are either dead or recovered at rate \(\gamma^i_{J,k}\).

\begin{equation} 
    \frac{dR_J I_k^i(t)}{dt} = \psi^i_{J, k} R_{J}E_{k}^i(t) - \gamma^i_{J,k} R_{J}I_{k}^i(t). 
    \label{eq:3} 
\end{equation}

Forth, in Eq. (\ref{eq:4}), \(\frac{dD^i(t)}{dt}\) is the dynamic number of dead individuals over time. For each strain \(k\), and for each group \(J \backslash \{k\}\), infected individuals who do not recover die at rate \(\gamma^i_{J \backslash \{k\},k}\) with probability \((\xi_{J \backslash \{k\},k})\).

\begin{equation} 
    \frac{dD^i(t)}{dt} = \sum_{k \in M, J \in P(M)} \gamma^i_{J \backslash \{k\}, k} \xi^i_{J \backslash \{k\},k} R_{J \backslash \{k\}}I_{k}^i(t).
     \label{eq:4}
\end{equation}

In summary, the single specie-level epidemiological dynamics take the form:
\begin{equation}
    \begin{array}{l}
    
    \frac{dR^i_J(t)}{dt} = \sum_{l \in J} \gamma^i_{J \backslash \{l\},l} (1-\xi^i_{J \backslash \{i\}, i}) R^i_{J \backslash \{i\}} I_{i}(t) - \sum_{l \in M_i \backslash J} \beta^i_{J, l} R^i_J(t) \sum_{L \in P(M) , c \not\in L} R_{L}I_{c}^i(t) , \\\\
    
    \frac{dR_J E_k^i(t)}{dt} = \sum_{k \in M_i \backslash J} \beta^i_{J,i} R_J^i(t) \sum_{L \in P(M_i) , k \not\in L} R_{L}I_{k}^i(t) - \psi^i_{J, k} R_{J}E_{k}^i(t), \\\\
    
      \frac{dR_J I_k^i(t)}{dt} = \psi^i_{J, k} R_{J}E_{k}^i(t) - \gamma^i_{J,k} R_{J}I_{k}^i(t), \\\\
    
    \frac{dD^i(t)}{dt} = \sum_{k \in M, J \in P(M)} \gamma^i_{J \backslash \{k\}, k} \xi^i_{J \backslash \{k\},k} R_{J \backslash \{k\}}I_{k}^i(t).. \\\\
        \label{eq:pandemic_final}
    \end{array}
\end{equation}
One can notice that the last equation that captures the number of individuals in the population that die due to the pandemic does not change the dynamics. As such, for our subsequent implementation, we will omit this equation from consideration. 

\subsection{Cross-Species dynamics}
\label{section:cross-species}
In our model, the cross-species dynamics include two main components: cross-infection and prey-predator interactions. However, not all species interact with all other species and, even if they do interact, they need not necessarily interact in the same way. As such, one can represent the interactions between a set of species \(\mathbf{P} := [\mathbb{P}_1, \dots, \mathbb{P}_N]\) using a directed, non-empty graph \(G := (V, E_1, E_2)\) where \(V \in \mathbf{P} \times \mathbb{R}^2\) is a set of nodes corresponding to the species populations, \(E_1 \subset V \times V \times \mathbb{R}^2 \) is set of directed edges representing the prey-predator interactions, and \(E_2 \subset V \times V \times \mathbb{R}^{|M_x||M_y|}\) is set of directed edges representing the cross-infection interactions. Formally, \(v \in V\) represents the entire population of specie with two parameters: the natural growth rate due to free resources \(a^i \in \mathbb{R}\) and natural population decay \(d^i \in \mathbb{R}\). The prey-predator interaction between specie \(\mathbb{P}_x\) and  \(\mathbb{P}_y\), \(e^{x,y}_1 \in E_1\), defines two parameters the average portion of population \(\mathbb{P}_x\) consumes from population \(\mathbb{P}_y\), \(C_{x,y} \in \mathbb{R}\), and the growth rate population \(\mathbb{P}_x\) obtains from consuming population \(\mathbb{P}_y\), \(B_{x,y} \in \mathbb{R}\). The cross-infection interaction between specie \(\mathbb{P}_x\) and \(\mathbb{P}_y\), \(e^{x,y}_2 \in E_2\), defines the average infection rate from an infected individual with strain \(k_x\) that belongs to population \(x\) to an individual in population \(y\) to become exposed to strain \(k_y\) to be \(\beta^{x,y}_{k_x, k_y} \in \mathbb{R}\). 

Accordingly, the prey-predator dynamics is following the Lotka-Volterra model for two species at a time. As such, in Eq. (\ref{eq:5}), \( \frac{d|\mathbb{P}_x(t)|(t)}{dt}\) is the dynamic number of individuals in population \(x\) over time. It is influenced by the following four terms. First, the non-infected population has a natural reproduction at rate \(a_x\). Second, the \(x\) population is consuming a set of other populations, \(\{y | (x, y) \in E_1 \}\), such that from each one of them with a rate \(B_{x,y}\) is added to the \(x\) population concerning the size of the \(y\) population. Third, in a symmetric way to the second term, the \(x\) population is also consumed by other species with a rate \(C_{x,y}\). Finally, the population's size is naturally exponentially decreased at a rate \(d_x\). 

\begin{equation} 
    \frac{d|\mathbb{P}_x(t)|}{dt} = a_x \sum_{J \in P(M_x)} R^x_{J}(t) + \sum_{\{y | (x, y) \in E_1 \}} B_{x, y}|\mathbb{P}_x(t)||\mathbb{P}_y(t)| -  \sum_{\{y | (y, x) \in E_1 \}} C_{y, x}|\mathbb{P}_y(t)||\mathbb{P}_x(t)| - d_x |\mathbb{P}_i(t)|.
     \label{eq:5}
\end{equation}

In addition, in order to capture the cross-infection dynamic, let us focus on two species, \(x\) and \(y\). For any two species, a matrix \(A \in \mathbb{R}^{|M_x||M_y|}\) is defined to represent the infection rate that makes individuals from the \(x\) population infected by strain \(k_1 \in M_x\) to an individual in the \(y\) population that would be infected by strain \(k_2 \in M_y\). Once \(x,y,k_1\) and \(k_2\) are chosen, the dynamic change of the exposed individuals in strain \(k_1\) from population \(x\) is corresponding to the infection rate \(\beta^{x,y}_{k_1, k_2}\) of all individuals in the \(x\) population that are susceptible to strain \(k_1\) and can be infected from an individual from population \(y\) that is infected by strain \(k_2\) with any recovery history, as formally described in Eq. (\ref{eq:6}):

\begin{equation} 
   \frac{dR_JE_{k_1}^x(t)}{dt} = \beta^{x,y}_{k_1, k_2} \sum_{J \in M_x \backslash \{k_1\}} dR_J^x(t) \sum_{L \in M_y \backslash \{k_2\}} R_LI_{k_2}^y(t).
     \label{eq:6}
\end{equation}

Hence, these dynamics takes can be summarized by the following system of ODEs:
\begin{equation}
    \begin{array}{l}
    
    \forall x \in V: \frac{d|\mathbb{P}_x(t)|(t)}{dt} = a_x \sum_{J \in P(M_x)} R^x_{J}(t) + \sum_{\{y | (x, y) \in E_1 \}} B_{x, y}|\mathbb{P}_x(t)||\mathbb{P}_y(t)| -  \sum_{\{y | (y, x) \in E_1 \}} C_{y, x}|\mathbb{P}_y(t)||\mathbb{P}_x(t)| - d_x |\mathbb{P}_i(t)|, \\\\
    
    \forall (x, y) \in E_2, k_1, k_2 \in M_x \times M_y: \frac{dR_JE_{k_1}^x(t)}{dt} = \beta^{x,y}_{k_1, k_2} \sum_{J \in M_x \backslash \{k_1\}} dR_J^x(t) \sum_{L \in M_y \backslash \{k_2\}} R_LI_{k_2}^y(t) 
    
        \label{eq:prey_predator_final}
    \end{array}
\end{equation}
where \(|\mathbb{P}_i(t)| := \sum_{k \in M} \sum_{J \in P(M \backslash \{k\})} \big (R_J I_k^i(t) + R_J E_k^i(t) \big ) + \sum_{J \in P(M)} R_J(t)\) is the \(i_{th}\) population's size at time \(t\). 

\begin{figure}[!ht]
    \centering
    \includegraphics[width=0.5\textwidth]{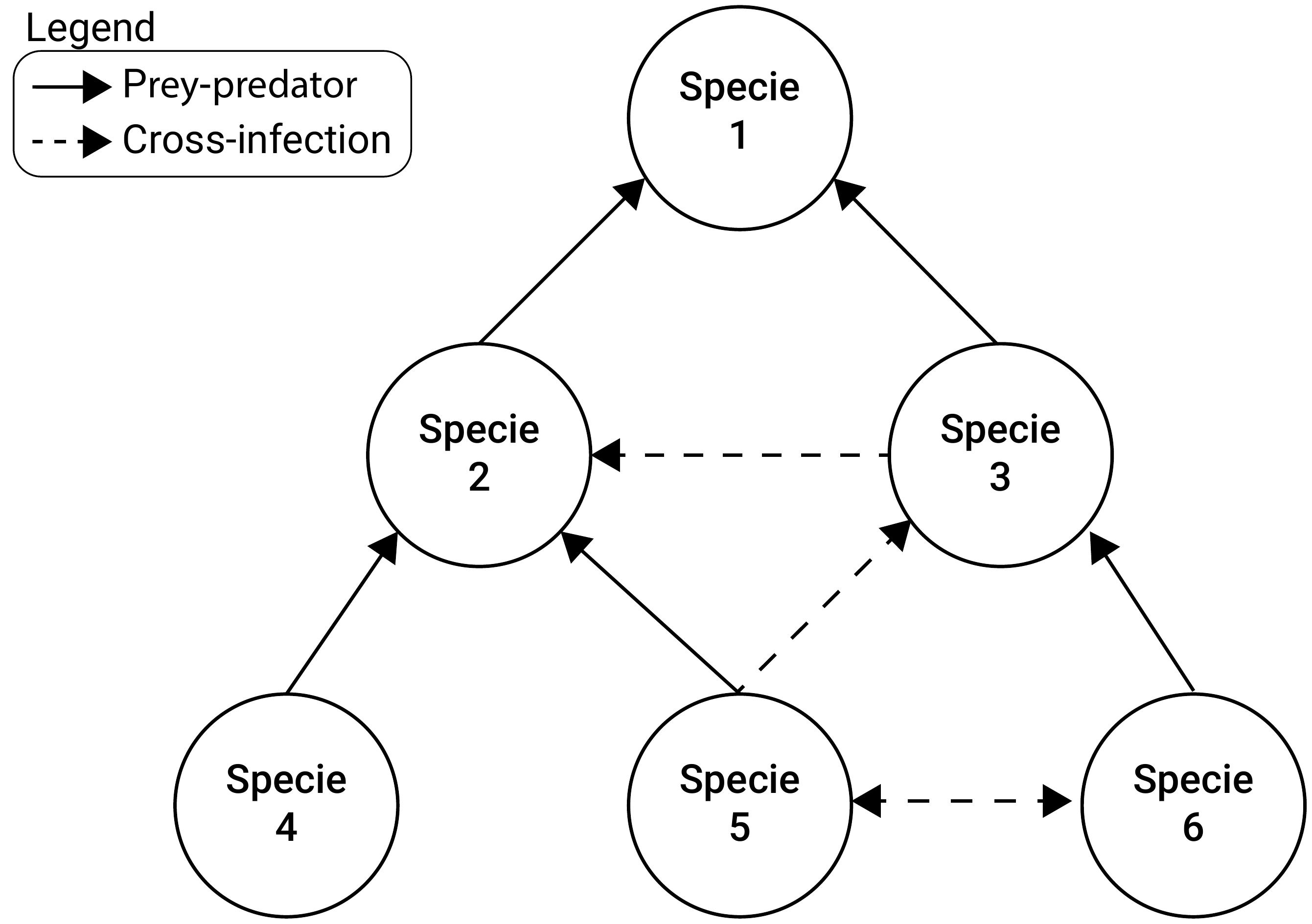}
    \caption{An example for prey-predator and cross-infection graph of multi-species with six species.}
    \label{fig:pp_scheme}
\end{figure}

A schematic example of the prey-predator and cross-infection in a multi-species case is provided in Fig. \ref{fig:pp_scheme} where six species are participating in the dynamics such that specie 1 eats species 2 and 3, specie 2 eats species 4 and 5, and specie 3 eats specie 6. In addition, specie 3 infects specie 2 and is infected by specie 5. Species 5 and 6 infect each other.  

\section{Computer Simulation}
\label{sec:simulator}
To simulate the model, we used an agent-based simulation approach \cite{abs_1,abs_2,teddy_pandemic_management,abs_3,abs_4} where each individual in the multi-population (i.e., the set of all species populations) is a timed finite state machine \cite{fsm} that defined by a tuple \(p_i \in \mathbb{P}_i: p_i := (s, k, J, \tau)\) such as \(s \in [1, \dots, N]\) is the individual's specie's index, \(k \in M_i \cup 0\) is the current strain the individual is infected with (if any), \(J \in P(M_i)\) is the recovery history, and \(\tau \in \mathbb{N}\) is the number of time steps that passed from the last change in the epidemiological state.

At the beginning of the simulation, the user is responsible to generate a set of populations, such that all agents in the same population have an identical \(s\) value. Moreover, for each population \(i\), the user declares the number of strains \(M_i\) and as a result, provides the following set of parameters: infection rates \(\beta_{J, k}^i\), duration from exposure to being infectious \(\psi_{J, k}^i\), recovery duration \(\gamma_{J, k}^i\), and recovery rate \(\xi_{J, k}^i\), for each \(k \in M_i\) and \(J \in P(M_i)\). In addition, for each population \(i\), the user provides natural growth and decay rates \(a_i\) and \(d_i\), respectively. Once all the populations are generated, the user is required to construct the prey-predator graph \(G\) by introducing the two sets of edges \(E_1\) and \(E_2\) as follows. First, for each pair of populations such that population \(x\) is the prey and population \(y\) is the predator, the edge \((x, y) \in E_1\) is added with the consumption portion of the prey population \(B_{x,y}\) and the growth to the predator population \(C_{x,y}\). Second, for each pair of populations \(x\) and \(y\) (not necessarily prey and predator), the edge \((x, y) \in E_2\) is added such that the \(x\)'s individual's strain \(k_1 \in M_x\), the \(y\)'s individual's strain \(k_2 \in M_y\) with recovery history \(J \in P(M_y)\), and an cross-infection rate \(\beta^{x,y}_{J,k_1} \in \mathbb{R}^+\).

\begin{figure}[!ht]
    \centering
    \includegraphics[width=0.99\textwidth]{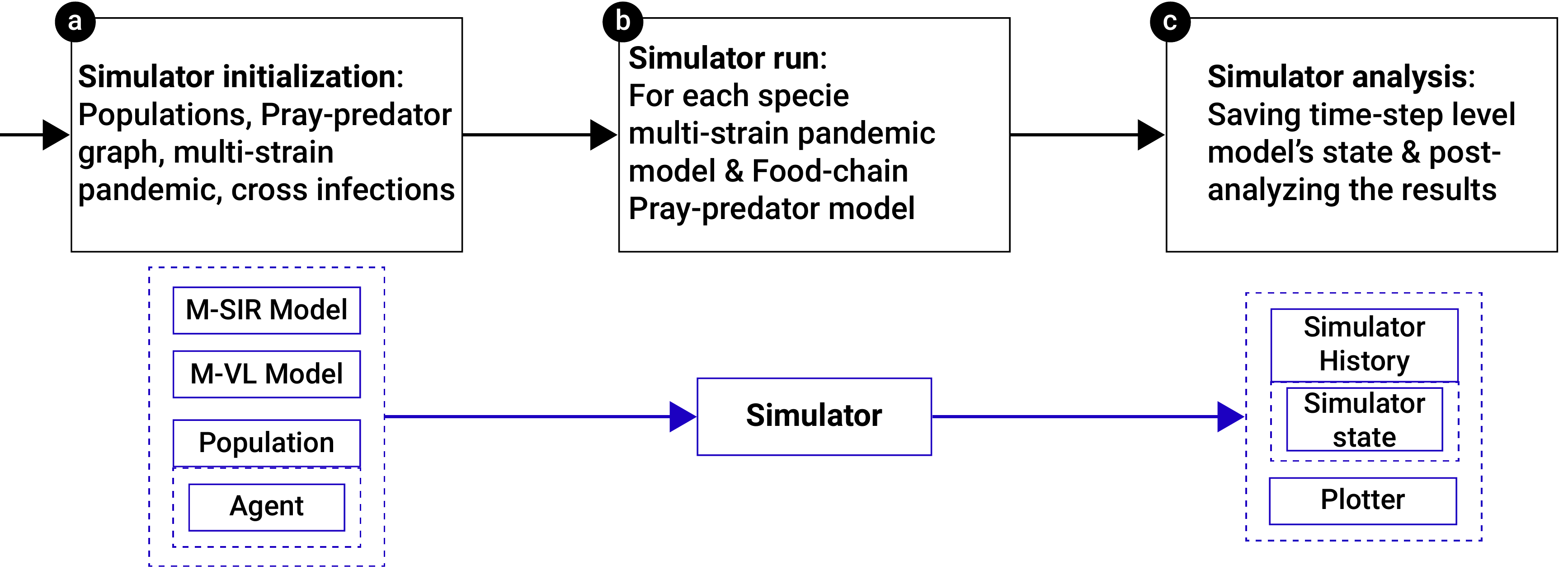}
    \caption{A schematic view of the simulator's process in both process diagram (black) and objects diagram (blue). (a) the initial phase of constructing the simulator, (b) the computation of the simulation itself, and (c) the analysis of the simulation's results.}
    \label{fig:simulator}
\end{figure}

Afterward, the simulation takes place in rounds \(r \in [1, \dots, T]\) such that \(T < \infty\). At each round, individuals in each population may interact, thus initiating some epidemiological and prey-predator dynamics as mathematically detailed in sections \ref{section:inner-species} and \ref{section:cross-species}. However, since the order of execution of each dynamic might influence the course of the dynamics, we tackle this challenge by computing all the changes in the meta-population and executing all of them at once after canceling-out opposite changes, as commonly performed in particle simulations \cite{particle_sim_1,particle_sim_2,particle_sim_3}. A schematic view of the simulator's process is shown in Fig.~\ref{fig:simulator}.

\subsection*{Ecosystem instability metric}
There are multiple metrics used to evaluate the course of a pandemic, such as total mortality, the maximum number of infected individuals at the same time, and the basic reproduction number \cite{metric_paper_1,metric_paper_2,teddy_multi_strain,metric_paper_4,metric_paper_3}. Each of these properties captures different properties of the pandemic spread. However, they are all designed for the case where there is a single population or where the evaluation is agnostic to all sub-populations and aims to measure the overall pandemic spread. 

Thus, for our analysis, we propose a novel metric to measure the pandemic spread for a multi-species scenario where the focus is on the ecosystem's ability to reach back a stable state. Intuitively, if no pandemic is present, the ecosystem reaches an equilibrium state \cite{pp_equlibrium_1,pp_equlibrium_2,pp_equlibrium_3}, \(s^*\), that can be treated as a baseline and therefore \(d(s^*) = 1\). The other extreme case, \(s^{**}\) is that all species are extinct due to the pandemic. As this is probably the worst-case scenario, we define  \(d(s^{**}) = 0\). Notice that both \(s^*\) and \(s^{**}\) define an equilibrium state. Thus, it is self-evident to require a metric \(d\) to evaluate the system's closest equilibrium state's condition. Following this rationale, the metric \(d\) is assessing an equilibrium state \(s\) that indicates the number of extinct species. Moreover, as we allow the equilibrium state to contain any value in \(\mathbb{R} \cup {-\infty, \infty}\), the proposed metric is also an indicator of the instability caused to the ecosystem by the pandemic. Therefore, the metric \(d\) is formally defined as follows.

\begin{definition}[Ecosystem's stability metric]
Given an MSMS dynamic system \(M\) with \(N\) species, the ecosystem's stability metric, \(d\), measures the level of ecosystem's stability significantly after the pandemic has been eliminated or has stabilized as follows:
\[d(M) := 1 - \frac{|\{v \in \lim_{t \rightarrow \infty} M(t): v \in \{0, \infty\} \}|}{N}\]
\end{definition}

Notably, one can define \(d\) slightly differently following the same motivation and constraints. Thus, the proposed definition is a sample of a feasible definition for \(d\) and not the only possible one. 

\section{Evaluation}
\label{sec:analysis_numerical}
In order to study the behavior of the MSMS model for various conditions and scenarios, we divide the analysis into two main parts: synthetic and real-world setups. Using the synthetic data, we are able to numerically study the damage and instability a multi-strain pandemic causes to an ecosystem for different cases. In particular, as the ecosystems are widely diverse and complex, in order to study the sensitivity of the pandemic's damage, we randomly generate a large set of cases and measure the average and standard deviation of the pandemic's damage on this set while changing a sub-set of parameters and initial conditions each time. In a complementary manner, real-world cases are considered in order to test the influence of global pandemic properties on the overall ecosystem's stability, given a specific topology of the species interactions. 

\subsection{Synthetic Setup}
For realizing this simulation, several parameters of the MSMS model have to be set. We discuss the main parameter values below and provide a summary in Table \ref{table:models_parameters}. The parameters are chosen to represent relatively small, yet diverse, ecosystems that require reasonable computational burden to simulate. The pandemic's properties as well as the prey-predator dynamics are randomly sampled, if not stated otherwise. 

\begin{table}[!ht]
\centering
\begin{tabular}{|p{0.1\textwidth}|p{0.6\textwidth}|p{0.2\textwidth}|}
\hline
\textbf{Parameter} & \textbf{Description} & \textbf{Default value} \\ \hline
\(|V|\) & Number of species [1] & \([5, \dots, 20]\) \\ \hline
\(|P_i|\) & Number of individuals in the \(i_{th}\) species' population [1] & \([500, \dots, 5000]\) \\ \hline
\(|E_1|\) & Number of prey-predator connections [1] & \( [0.05|V|,  \dots, 0.5|V|] \) \\ \hline
\(|E_2|\) & Number of cross-infection connections [1] & \( [0.05|V|,  \dots, 0.5|V|] \) \\ \hline
\(|M_i|\) & Number of strains for the \(i_{th}\) species [1] & \( [0, 1, 2, 3, 4] \) \\ \hline
\(T\) & Steps in time in days [\(t\)] & \( 365 \) \\ \hline
\(\beta_{k, J}^i\) & The infection rate of specie \(i\) for strain \(k\) for an individual with recovery history \(J\) [1] & \( [0.05, \dots, 0.15] - [0.01, \dots, 0.1] \cdot |J|/|M_i| \) \\ \hline
\(X_{k, J}^i\) & The strain \(X \in [\gamma, \phi, \xi, \psi]\) property rate of specie \(i\) for strain \(k\) for an individual with recovery history \(J\) [1] & \( [0.01, \dots, 0.5] \) \\ \hline
\end{tabular}
\caption{The model's parameters description and default values.}
\label{table:models_parameters}
\end{table} 

\begin{figure}[!ht]
    \begin{subfigure}{.5\textwidth}
        \includegraphics[width=0.99\textwidth]{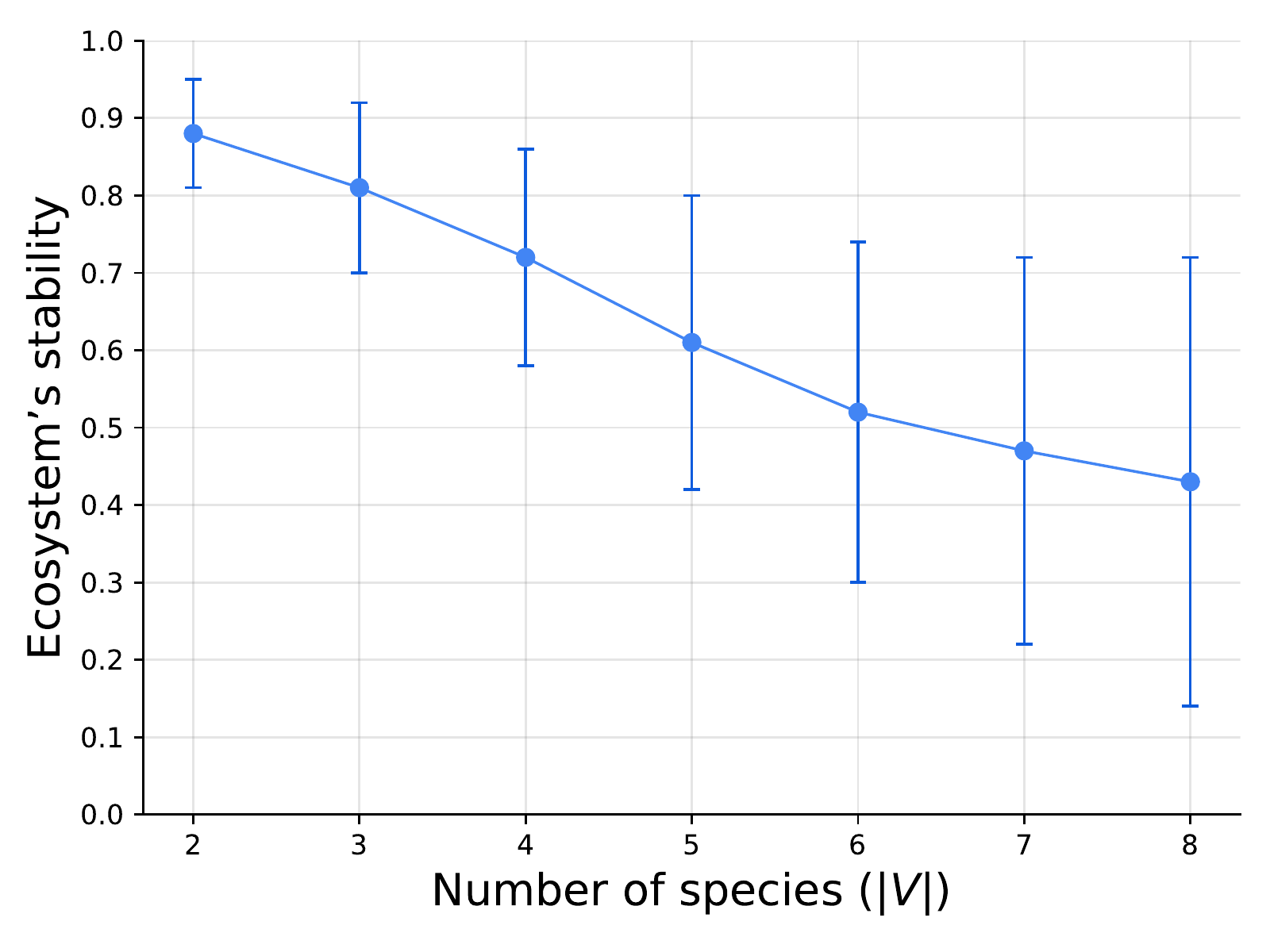}
        \caption{Number of species (\(|V|\)).}
        \label{fig:sens_species_countt}
    \end{subfigure}
    \begin{subfigure}{.5\textwidth}
        \includegraphics[width=0.99\textwidth]{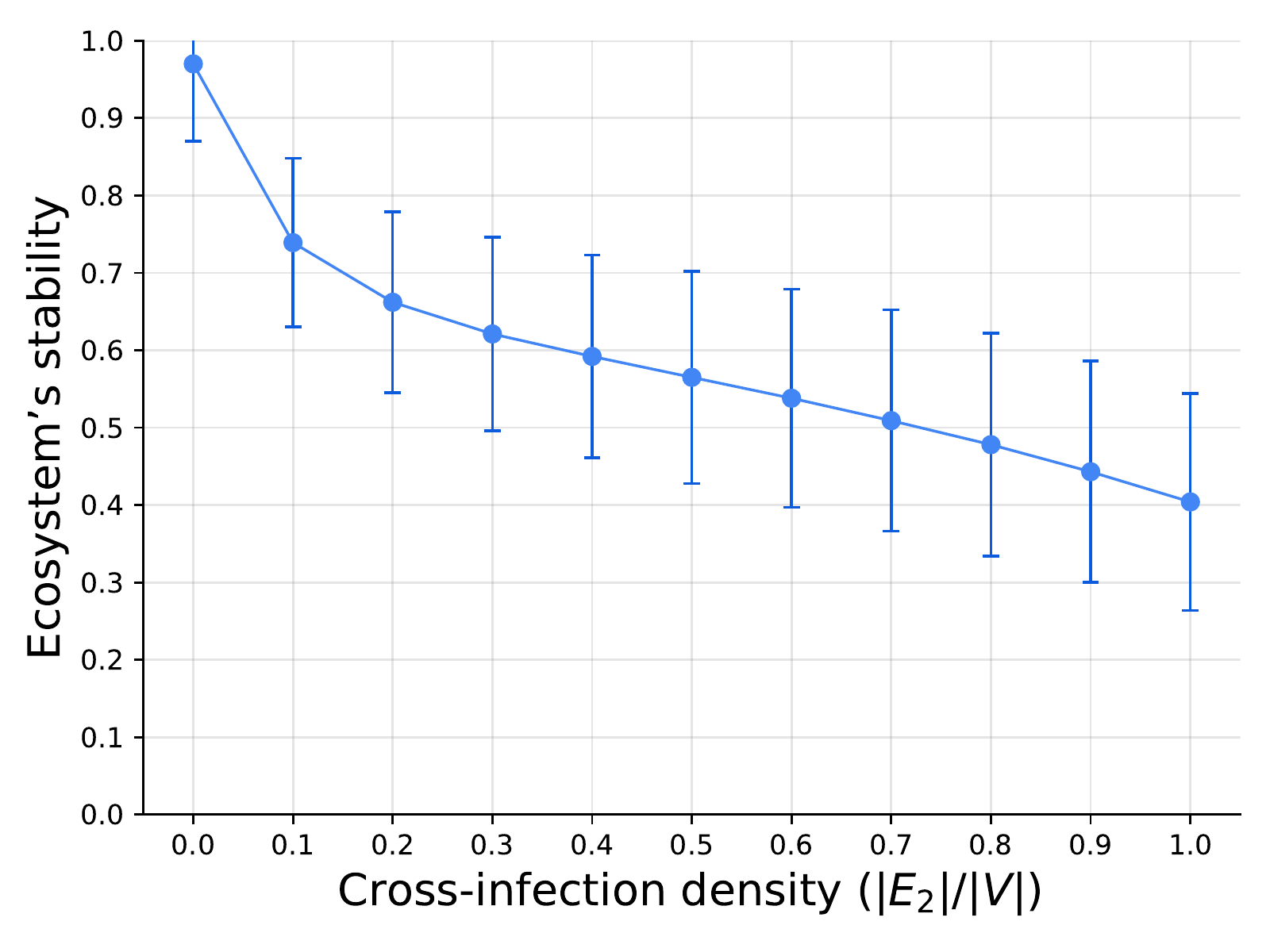}
        \caption{Cross infection density (\(|E_2|/|V|\)).}
        \label{fig:sens_ci_density}
    \end{subfigure}
    
    \begin{subfigure}{.5\textwidth}
        \includegraphics[width=0.99\textwidth]{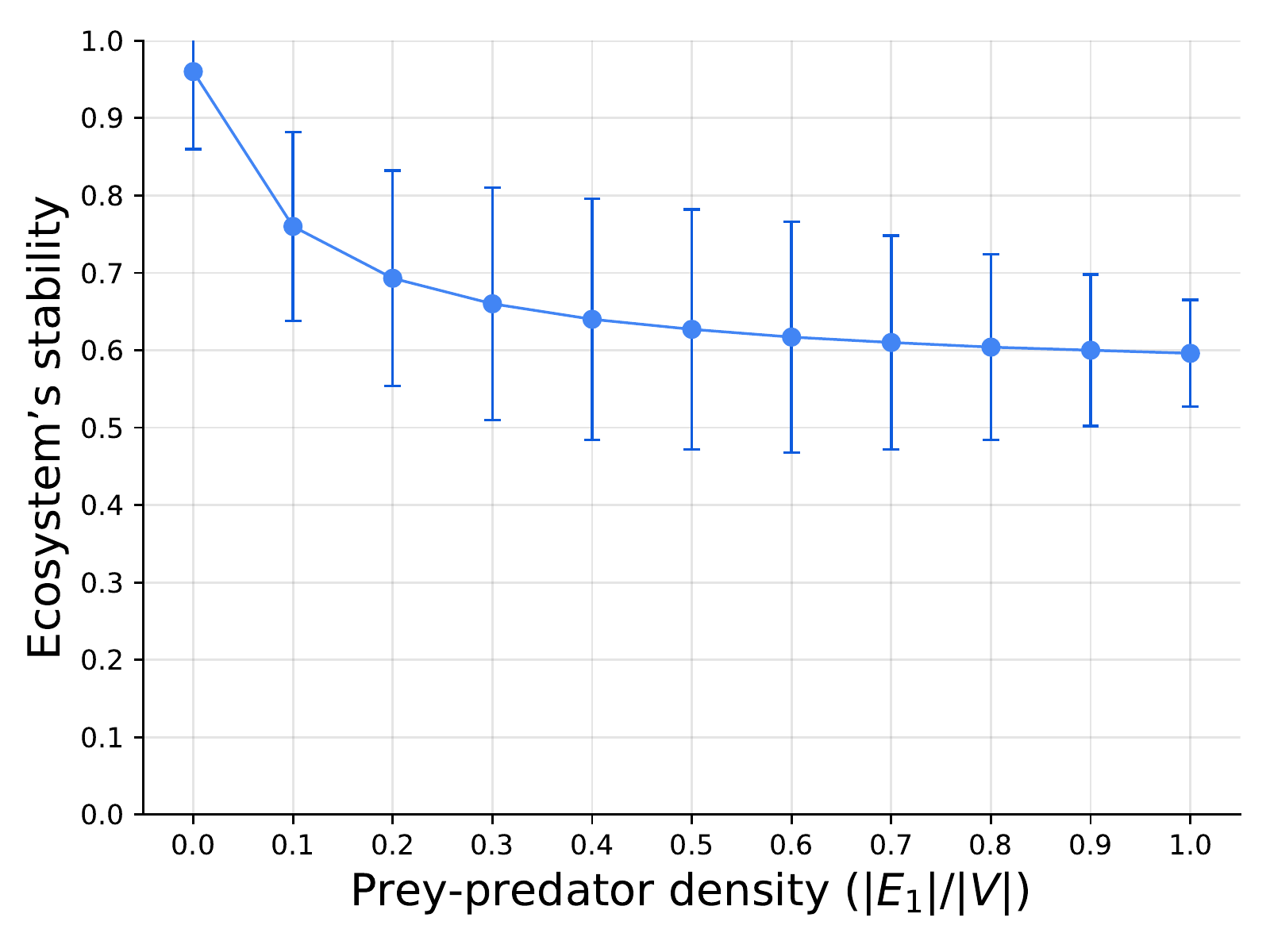}
        \caption{Prey-Predator density (\(|E_1|/|V|\)).}
        \label{fig:sens_pp_density}
    \end{subfigure}
    \begin{subfigure}{.5\textwidth}
        \includegraphics[width=0.99\textwidth]{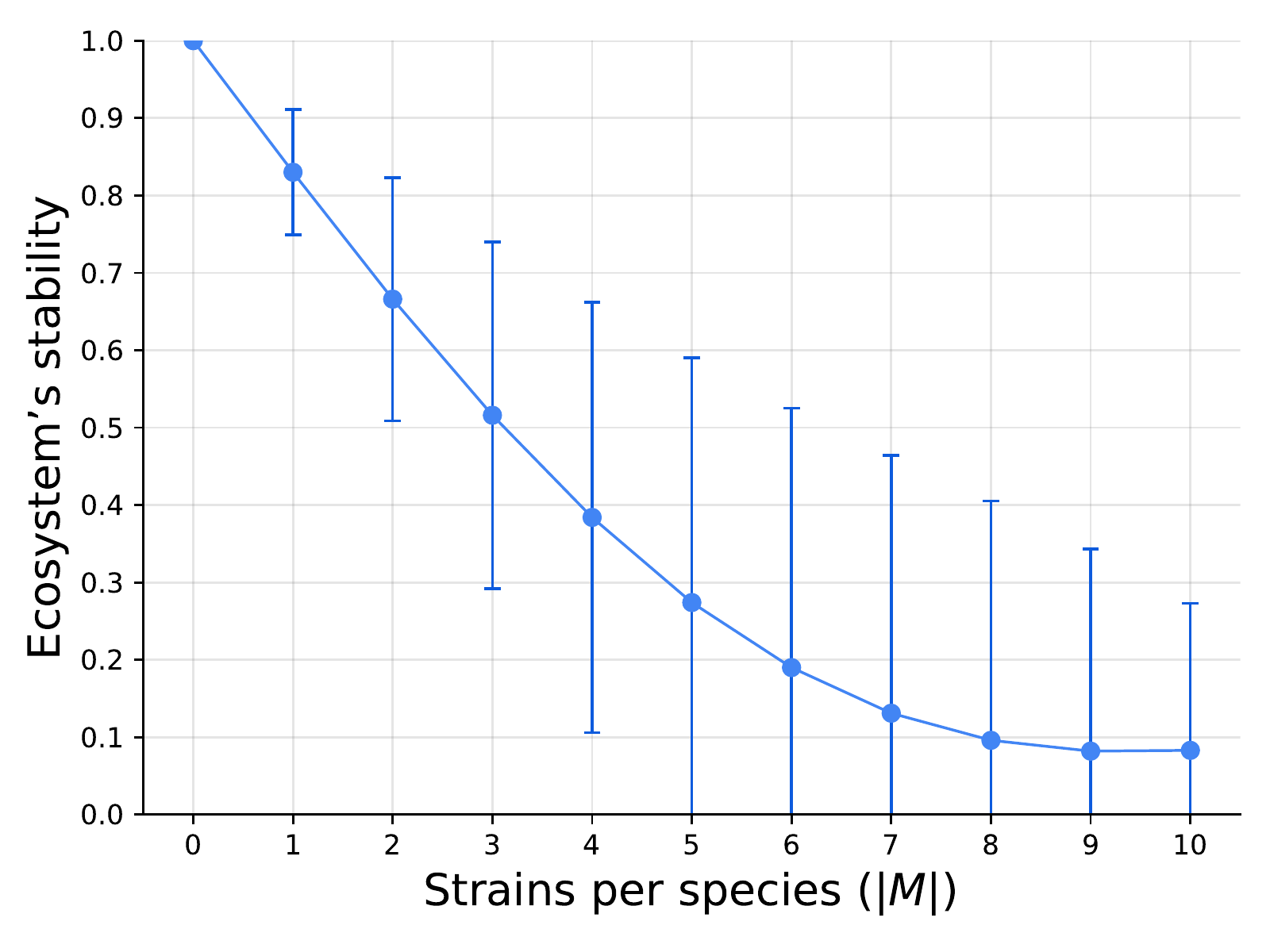}
        \caption{Number of strains (\(|M|\)).}
        \label{fig:sens_strains}
    \end{subfigure}
    
    \caption{A sensitivity analysis of the ecosystem's stability metric. The results are shown as mean \(\pm\) standard deviation of \(n = 100\) repetitions.}
    \label{fig:sensitivity_all}
\end{figure}

In order to obtain a wide variety of cases, \(n=1000\) randomly generated graphs are considered. For the sampled 1000 graphs, we examine the ecosystem's stability metric as a function of different properties of the MSMS model. Fig. \ref{fig:sensitivity_all} summarizes the main results obtained. As one can see from Fig. \ref{fig:sens_species_countt}, the mean ecosystem's stability is monotonically decreasing by the number of species (\(|V|\)) and the standard deviation is increasing. In a similar manner, Fig. \ref{fig:sens_ci_density} shows that the cross-infection density  (\(|E_2|/|V|\)) enforces a monotonic decreasing behavior to the ecosystem's stability. Moreover, as the cross-infection density increases, the decrease rate intensifies, as indicated by the negative value of the second-order numerical derivative of the graph. The ecosystem's stability shows a monotonically decreasing behavior to the prey-predator density (\(|E_1|/|V|\)), as presented by Fig. \ref{fig:sens_pp_density}. Specifically, an inverse behavior is found to be \(0.95 - 0.39Z/(Z+0.11)\) where \(Z := |E_1|/|V|\) with a coefficient of determination \(R^2 = 0.958\), using the least mean square method \cite{lma}. Similar dynamics are encountered when computing the sensitivity of the ecosystem's stability to the number of strains (\(|M|\)), as shown in Fig. \ref{fig:sens_strains}. 

\subsection{Real-World Setup}
\label{sec:real_world}
In order to evaluate the proposed MSMS model's ability to capture and predict the ecosystem's state during a pandemic in a more realistic setup, we consider two real-world cases: First, a wild farm case in Asia where farm animals have interacted with nearby wild animals; Second, a near-shore ocean case. For both cases, we consider the Avian Influenza virus to be the pathogen at the root of the pandemic \cite{virus_1,virus_2,virus_3,virus_4,virus_5,virus_6}. 

\begin{figure}[!ht]
    \begin{subfigure}{.5\textwidth}
        \includegraphics[width=0.9\textwidth]{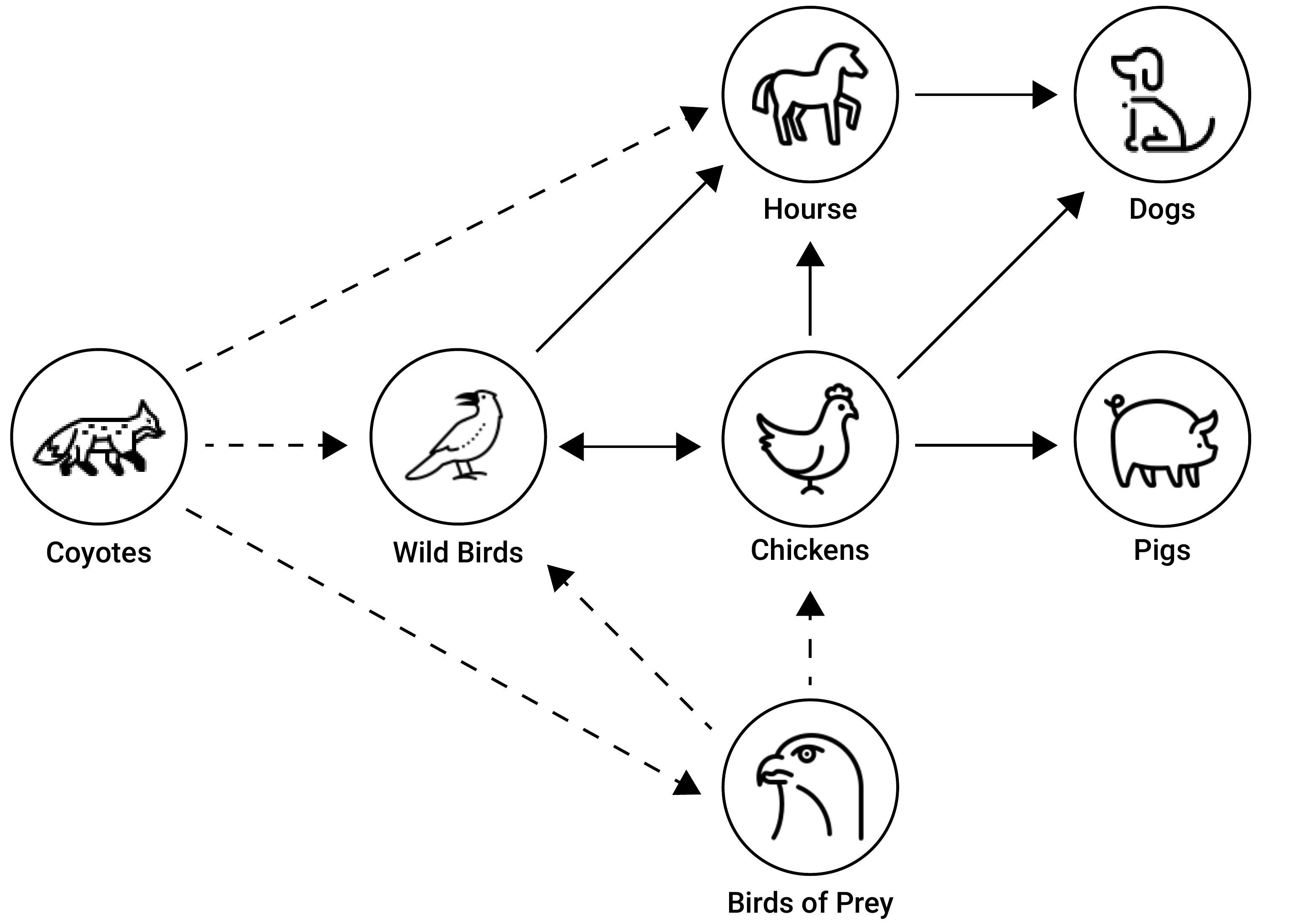}
        \caption{Wild farm settings.}
        \label{fig:real_wolrd_process_1}
    \end{subfigure}
    \begin{subfigure}{.5\textwidth}
        \includegraphics[width=0.9\textwidth]{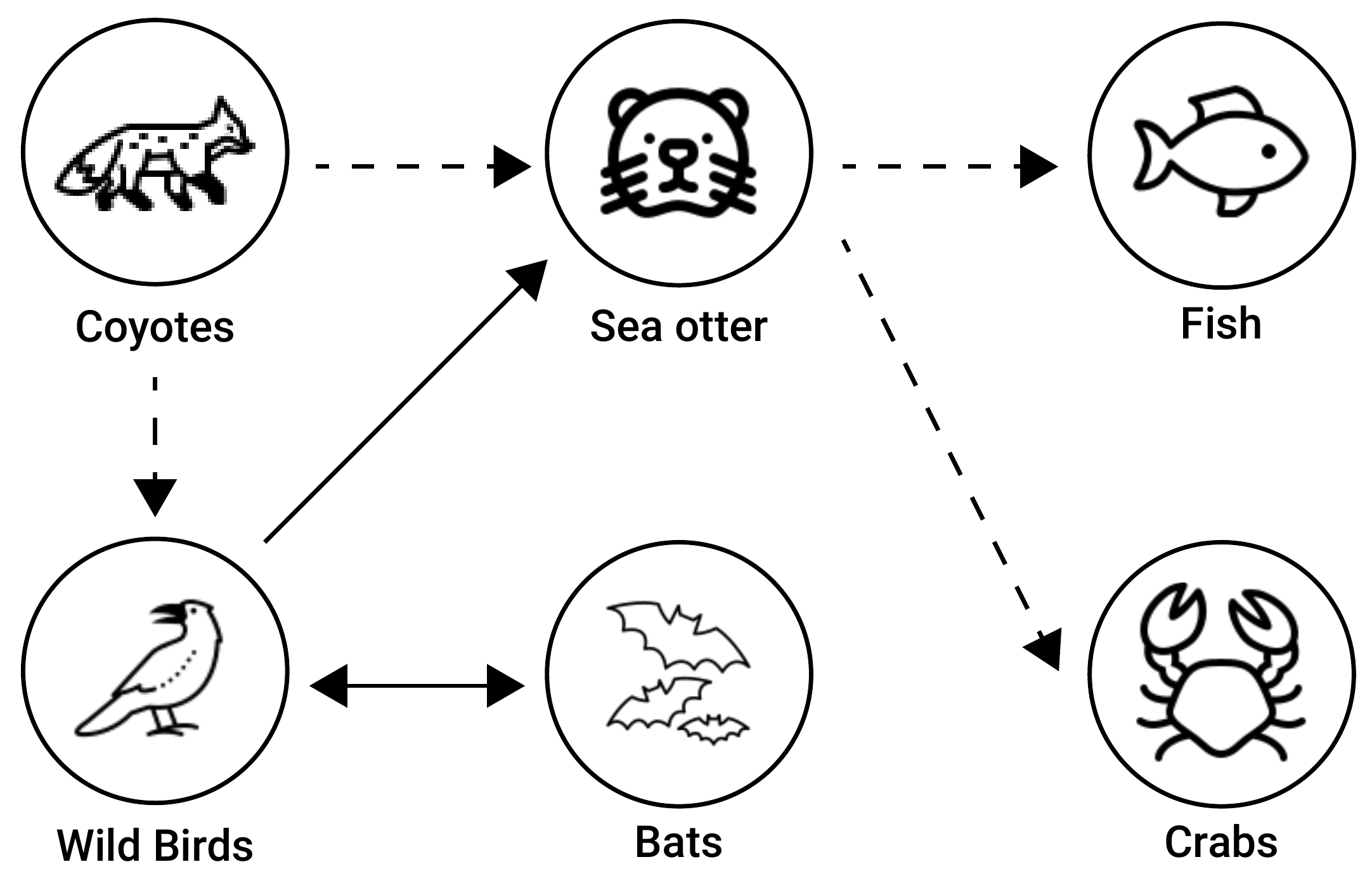}
        \caption{Near-shore ocean settings.}
        \label{fig:real_wolrd_process_2}
    \end{subfigure}
    
    \caption{A schematic view of the two real-world cases. The solid and dashed lines indicate the cross-infection and prey-predator interactions between species, respectively.}
    \label{fig:real_cases}
\end{figure}

Due to the fact that each case involving many species and their interactions, there is a lot of unknown information.
In order to maintain a balance between available data and computational resources and the case's representation accuracy, we have selected a subset of species that play a key role in the dynamics. For both cases we consider the following: Initially, wild birds infected farm chickens, then chickens infected wild birds. Furthermore, wild birds and chickens infect horses, which in turn infect dogs. In addition, chickens also directly infect dogs and pigs \cite{real_world_infection}. Among the prey-predator interactions, Coyotes eat wild birds, birds of prey, and horses \cite{real_world_eat_1}. In addition, birds of prey eat smaller wild birds \cite{real_world_eat_2}. In the second case, wild birds infect bats and vice versa. In addition, they infect sea otters \cite{real_world_infection}. As a prey-predator interaction, Coyotes eat wild birds and sea otters \cite{real_world_eat_1}. In their turn, sea otter eats small fish and crabs \cite{real_world_eat_3}. A schematic view of the two cases is shown in Fig. \ref{fig:real_cases}, where the solid and dashed lines indicate the cross-infection and prey-predator interactions, respectively. 

In order to use the proposed MSMS model in these cases, one first needs to find the model's parameters' values and initial condition. Unfortunately, this data is largely unavailable \cite{human_animal_science} and even partial observations of the dynamics highly differ between locations and timeframes \cite{eco_hard_1,eco_hard_2,eco_hard_3}. To overcome this challenge, we generate a large number of samples under the constraint that for the same initial condition and prey-predator interactions, the ecosystem's stability after 365 steps in time is 1. The motivation behind this constrain is to sample cases that are relatively stable if no pandemic is presented, as believed to be the case for a short duration of time in most setups \cite{lv_example_2,lv_example_3,lv_example_4}.

\begin{figure}[!ht]
    \begin{subfigure}{.5\textwidth}
        \includegraphics[width=0.99\textwidth]{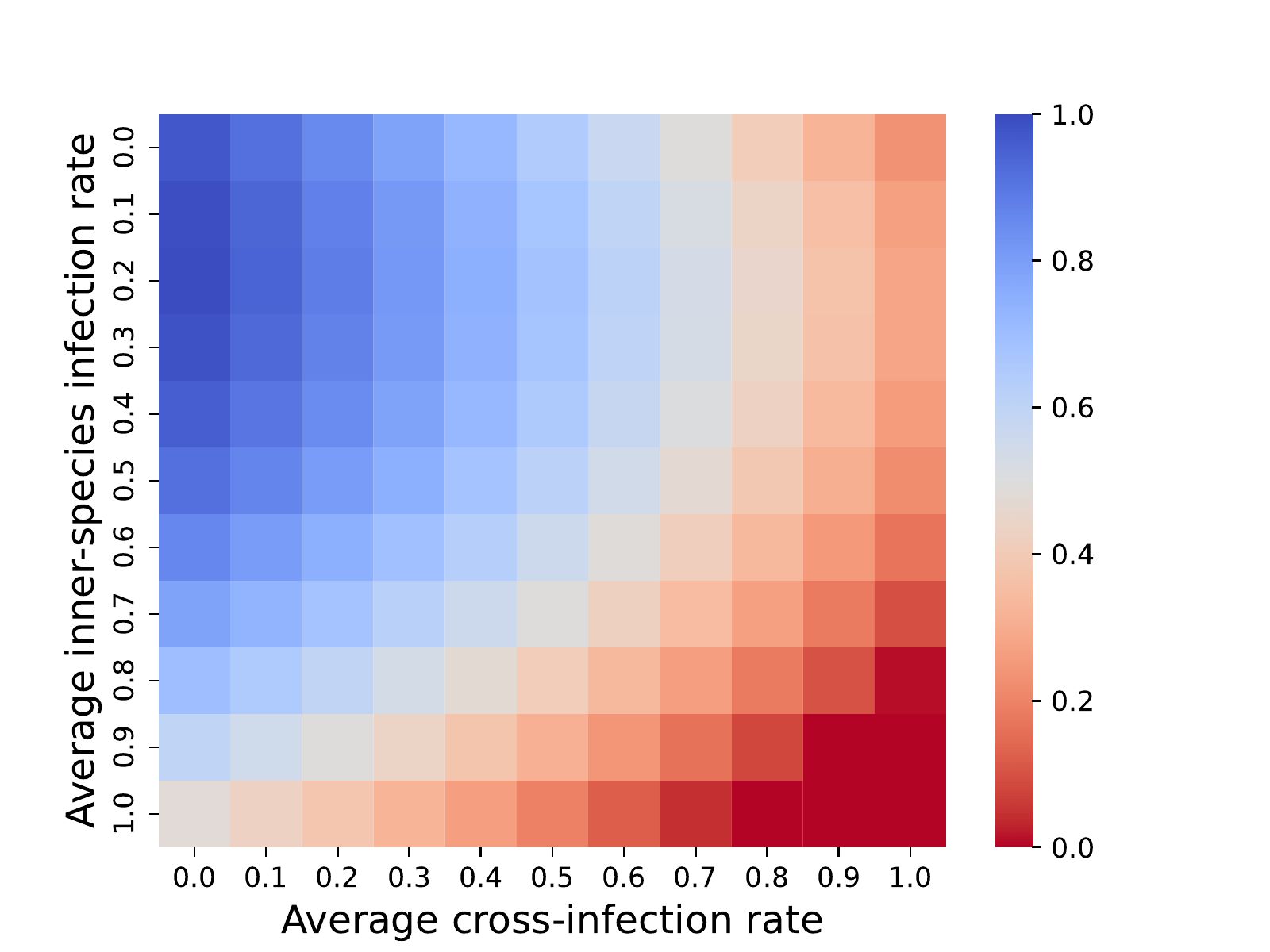}
        \caption{Wild farm case.}
        \label{fig:real_wolrd_process_1}
    \end{subfigure}
    \begin{subfigure}{.5\textwidth}
        \includegraphics[width=0.99\textwidth]{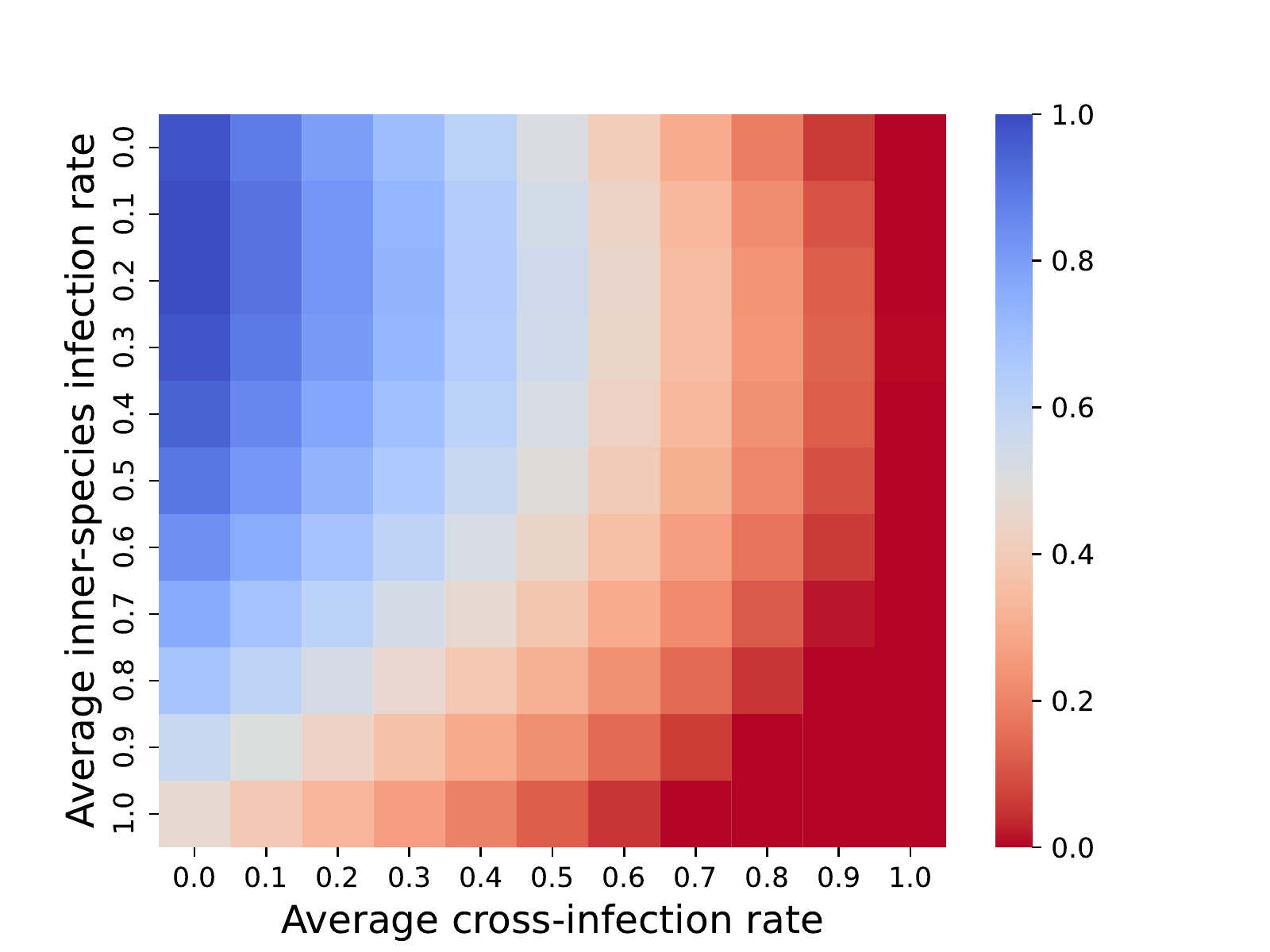}
        \caption{Near-shore ocean case.}
        \label{fig:real_wolrd_process_2}
    \end{subfigure}
    
    \caption{A two-dimensional sensitivity analysis for the real-world cases between the ecosystem's stability and the combined influence of the average cross-infection rate and the average inner-species infection rate.}
    \label{fig:real_wolrd_process}
\end{figure}

Fig. \ref{fig:real_wolrd_process} presents a two-dimensional sensitivity analysis for the real-world cases between the ecosystem's stability and the average cross-infection rate on the x-axis and the average inner-species infection rate on the y-axis. The results are shown as the average of \(n = 25\) samples for each configuration, as computed after \(T = 365\) steps. A paired T-test between the result metrics of the \say{wild farm} and \say{near-shore ocean} case shows that the two cases statistically significantly differ with \(p < 0.0001\). One can see in both cases that, on average, a larger average cross-infection rate and a larger inner-species infection rate cases more instability in the ecosystem. However, this connection is non-linear as linear fitting on both cases resulted in a coefficient of determinations \(R^2 = 0.312\) and \(R^2 = 0.185\), respectively. 

\section{Discussion}
\label{sec:discussion}
In this study, we proposed a novel ecological-epidemiological model of multi-species with multi-strain dynamics that account for prey-predator interactions and cross-infection interactions between arbitrary numbers of species using an extended Lotka-Volterra and SEIRD models, represented by ordinary differential equations operated on a graph. Considering the Avian Influenza A pathogen and its various strains for different species as a representative example, we evaluated the proposed model using an extensive agent-based simulation based on both synthetic and real-world graphs and data.

Starting with synthetic graphs and data, we examine the sensitivity of the ecosystem's long-duration stability due to the multi-strain pandemic, using a wide range of cases. On average, as the number of species increases, the ecosystem's stability decreases, as shown in Fig. \ref{fig:sens_species_countt}. Moreover, the entropy of the system is also increasing, as indicated by the standard deviation of the graph. These results agree with biological observations in nature \cite{oyster_density_eco_affect}. A similar outcome was achieved for cross-infection density and prey-predator density, as shown in figs \ref{fig:sens_ci_density} and \ref{fig:sens_pp_density}, respectively, aligning with the existing view that cross-infections in the predator population result in community instability among predators and their prey \cite{Stability_Prey_Behavioural_to_Predator_Density, oyster_density_eco_affect, stabilization_predator_prey, cross_infection_parasites}. Furthermore, as the number of strains in the environmental dynamics increases, the ecosystem's stability decreases. This outcome again agrees with prior literature examining other multi-strain pandemics applied to a single species \cite{teddy_multi_strain,multi_strain_2}. 

In addition, using real-world data, we examined two realistic cases - one of a wild farm and another of a near-shore ocean, presented in Fig. \ref{fig:real_cases}. We examined the influence of the average cross-infection rate and the average inner-species infection rate on the ecosystem's stability, as presented in Fig. \ref{fig:real_wolrd_process}. As one could expect, as these quantities increase, the ecosystem's stability decreases in a non-linear fashion. Comparing the cross-infection rates between wild farm animals, it seems that the ecosystem is more stable among them as opposed to those found near the shore. This observation also holds when we consider the rate of inner infection among these animals. Furthermore, both ecosystems appear to have a higher rate of cross-infection than within-species infection which makes the ecosystem less stable. 

Taken jointly, the results indicate that the proposed MSMS model with its agent-based simulation could adequately represent multi-species multi-strain pandemic dynamics. Researchers can utilize the proposed model to conduct \textit{in silico} experiments, exploring different pandemic intervention policies for a wide range of configurations.

This study has several limitations that should be addressed in future work to further improve the biological and ecological accuracy of the proposed MSMS model. First, as no spatial component is taken into consideration, the current infection rates are operating as an upper bound for the realistic infection rate \cite{teddy_multi_strain} which results in over-pessimistic outcomes. Moreover, by considering spatial dynamics, one is able to capture the movement patterns of various species \cite{spatial_1_example,spatial_2_example,spatial_3_example}. Thus, introducing a spatial component to the model would significantly increase its accuracy \cite{spatial_2,spatial_3,spatial_1}. Second, many species alter their behavior over time, due to wather chares for example, thus directly influencing other species' behaviors \cite{animal_behavior_1,animal_behavior_2,animal_behavior_3}. For instance, bird migration \cite{birds}, bears' hibernate \cite{bears}, and plants' blossom \cite{plants}. Third, the proposed model uses constant epidemiological and ecological values. However, in practice, these values are dynamic and influenced by the time of year, changes in the strains' mutation, and other factors. As such, time-depended or even stochastic values would make the model, presumably, even more realistic and accurate at the cost of analytical analysis feasibility. Finally, as strains in a pandemic are not static and new strains can appear through a mutation process in hosts, one can further extend the multi-strain pandemic model into a multi-mutation pandemic model as well \cite{teddy_labib_chaos}. 

\section*{Declarations}
\subsection*{Funding}
This research did not receive any specific grant from funding agencies in the public, commercial, or not-for-profit sectors.

\subsection*{Conflicts of interest/Competing interests}
The authors have no financial or proprietary interests in any material discussed in this article.

\subsection*{Data and materials availability}
The data that has been used is presented in the manuscript with the relevant sources.

\subsection*{Author Contributions}
Ariel Alexi: Conceptualization, investigation, data curation, and writing - original draft. \\ 
Ariel Rosenfeld: Conceptualization, supervision, validation, and writing - review \& editing. \\ 
Teddy Lazebnik: Conceptualization, formal analysis, methodology, software, data curation, methodology, project administration, visualization, supervision, and writing - original draft.

\subsection*{Acknowledgements}
The authors wish to thank Ziv Zemah Shamir for his ecological-related consulting.

\printbibliography

\end{document}